\documentclass[useAMS,usenatbib]{mn2e}
\usepackage{times}
\usepackage{graphicx, latexsym, amssymb, amscd, psfrag}
\usepackage{epsfig}

\title[Enhanced star formation on the outskirts of galaxy clusters]{Plunging
 fireworks: Why do infalling galaxies light up on the outskirts of clusters?}
\author[Mahajan, Raychaudhury and Pimbblet]{Smriti Mahajan\thanks{E-mail:
s.mahajan1@uq.edu.au}$^{1,2}$, Somak Raychaudhury$^{1}$ \& Kevin A. Pimbblet$^{3,4}$\\  
$^{1}$School of Physics and Astronomy, University of Birmingham, Birmingham B15~2TT, UK\\
$^{2}$School of Mathematics and Physics, University of Queensland, Brisbane, QLD 4072, Australia\\
$^{3}$Monash Centre for Astrophysics, Monash University, Clayton, VIC 3800, Australia\\
$^{4}$School of Physics, Monash University, Clayton, Melbourne, VIC 3800, Australia}

\newcommand{\s}{starburst}
\def\eg{{e.g. }}
\def\ie{{i.e. }}
\def\ssf{{SFR/$M^*$}}
\def\lx{{$L_X$}}

\begin{document}

\date{}

\pagerange{\pageref{firstpage}--\pageref{lastpage}} \pubyear{2010}
\maketitle

\label{firstpage}
%======================== ABSTRACT ==================================
\begin{abstract}

  Integrated star formation rate (SFR) and specific star formation
  rate (SFR/$M^*$), derived from the spectroscopic data obtained by
  SDSS DR4 are used to show that the star formation activity in
  galaxies ($M_{r} \leq -20.5$) found on the outskirts (1-2r$_{200}$)
  of some nearby clusters ($0.02\leq z \leq0.15$) is enhanced. By
  comparing the mean SFR of galaxies in a sample of clusters with
  at least one starburst galaxy (log SFR/$M^*\geq-10$ yr$^{-1}$ {\it{and}}
  SFR $\geq10$ M$_{\odot}$yr$^{-1}$) to a sample of 
  clusters without such galaxies (`comparison' clusters), we find that
  despite the expected decline in the mean SFR of galaxies toward the
  cluster core, the SFR profile
  of the two samples is different. Compared to the clusters
  with at least one starburst galaxy on their outskirts, the galaxies in
  the `comparison' clusters show a low mean SFR at all radius ($\leq
  3$r$_{200}$) from the cluster centre. Such an increase in the SFR
  of galaxies is more likely to be seen in dynamically unrelaxed 
 ($\sigma_v \gtrsim 500$ km s$^{-1}$) clusters. It
 is also evident that these unrelaxed clusters are currently being assembled
 via galaxies falling in through straight filaments, resulting in high velocity dispersions.
 On the other hand, `comparison' clusters are more likely to be fed by relatively low
 density filaments.    
  We find that the starburst galaxies on the periphery of clusters are 
  in an environment of higher local density than other cluster galaxies at
  similar radial distances from the cluster centre. We conclude that a relatively high
  galaxy density in the infalling regions of clusters promotes interactions
  amongst galaxies, leading to momentary bursts of star formation.
 Such interactions play a crucial role in exhausting the fuel for star
  formation in a galaxy, before it is expelled due to the
  environmental processes that are operational in the dense interiors
  of the cluster. 
  
\end{abstract}


%======================== Introduction===============================
\section{Introduction}

Statistical analyses of large samples of galaxies belonging to rich
galaxy clusters reveal a gradual trend of the quenching of star
formation in galaxies at decreasing radial distance from the centres
of clusters. This is generally interpreted as the overwhelming
influence of the gravitational field of the dark matter associated
with the cluster, added to the effect of the hot gas in the
intracluster medium (ICM), that serves to progressively deprive the
galaxies of the fuel for star formation through tidal and other
stripping processes. Thus, the core of a cluster is generally thought
to be the site of the most dramatic transformation in the evolution of
a galaxy. As a result, the central regions of galaxy clusters end up
being dominated by passively evolving red elliptical/lenticular
galaxies, while their blue star-forming counterparts prefer the
sparsely occupied regions away from clusters. 

 However, over the last few years, a growing number of galaxies, with
 an unusually high rate of star formation, have been discovered on the
 outskirts of rich galaxy clusters. These galaxies, often spotted
 close to or outside the virial radius of clusters, are seen to possess tidal
 features, and signs of morphological distortions that appear to be
 induced by close interactions. A majority of these galaxies have been
 discovered beyond a redshift of $z\!=\!0.2$
 \citep{keel05,moran05,sato06,marcillac07,oemler09}, since at these
 redshifts, the regions beyond the virial boundaries of clusters often
 fit on to current wide-format imagers. Good examples of these are the
 infalling galaxy C153 in Abell~2125 \citep[$z = 0.246$;][]{wang04},
 which exhibits comet-like tails of X-ray emitting gas, or the
 flagrantly star-forming galaxies on the outskirts of Abell~2744
 \citep[$z\!=\!0.4$;][]{braglia07}, which seem to lie on filaments
 feeding a set of merging clusters, and in the lensing cluster
 CL0024+16 \citep[$z\!=\!0.4$;][]{moran05}, where enhanced star
 formation is detected in a string of early-type dwarfs outside the
 virial radius.

 It is not yet known whether such galaxies are common at lower
 redshifts, since surveys away from the cores of nearby clusters are
 not common. Infalling galaxies like NGC~4472, at 1.2~Mpc from the
 core of the Virgo cluster ($z = 0.003$), show evidence of 
extended streams of hot X-ray emitting gas, and a shock front 
 in the X-ray observation \citep{kraft11}. This is accompanied by
 star
 formation in tidal streams of stripped cold gas from its dwarf
 companion UGC~7636 \citep{lee00}. Similar systems are seen elsewhere
 in Virgo \citep{abramson11}, and in Abell~3627 \citep[$z =
 0.016$;][]{sun10}, where the abundance of H$\alpha$ and HII emission
 indicate rapid star formation. Some of these galaxies are plunging
 towards the cluster core on their own, while others like those
 comprising J0454 ($z = 0.26$) are merging with the cluster as a
 galaxy group \citep{schirmer10}. In the absence of systematic
 studies, it is not clear how common such starburst galaxies are away
 from the cores of clusters, and indeed how important this phenomenon
 is in the grand scheme of galaxy evolution. One would like to know,
 for instance, whether their occurrence is related to the properties
 of the clusters, the structure of the surrounding cosmic web or the
 properties of the galaxies themselves.
 
 Earlier studies have suggested that evolution of galaxies in clusters
 is mainly driven by processes such as ram-pressure stripping, where a
 galaxy falling into cluster loses its gas content through interactions
 with the hot ICM, thus being deprived of star formation by the time it
 reaches the cluster core. Major mergers between galaxies are uncommon
 even in cluster cores \citep{lucia07}, and galaxy-galaxy interactions
  \citep{moore96} are in general thought to be
 unimportant in cosmological models \citep[e.g.][]{mayer06}. However, the
 observed radial gradient in star formation and other galaxy properties
 \citep[e.g.][]{boselli06,p08,mahajan10}, and the
 existence of transitional (i.e. post-starburst) galaxy populations to
 much farther distances from the centres of clusters than their region
 of influence \citep[e.g.][]{gavazzi10,mahajan11}, indicates that the
 influence of the cluster does not tell the whole story.
 
 In particular, repeated high velocity encounters between galaxies,
 often termed as ``harassment" \citep{moore96}, can play a special
 role in the evolution of galaxies in low and intermediate density
 environments \citep[][among
 others]{lewis,h2,p2,moss06,p08,oemler09,smith10}. The frequency and
 impact of such galaxy-galaxy interaction processes depend upon the
 local galaxy density.  So, in principle, any location on the cosmic
 web, with galaxy density above a `critical' value \citep[see][for
 instance]{g3}, should encourage interactions, resulting in bursts of
 star formation, among galaxies.  It has been shown elsewhere that
 filaments of galaxies \citep{p2,boue08,fadda,p08,edwards10,b11} and
 outskirts of clusters \citep{r2,smith10,coppin11,v12} are such
 favourable sites.

 While simulations suggest that stripped HI and X-ray emitting gas
 should be ubiquitous and detectable in the central regions of most
 clusters \citep[e.g.][]{tonnesen10,yamagami11}, it is difficult to
 quantify systematic properties of galaxies residing on inter-cluster
 filaments on the cosmic web due to the small numbers involved on a
 single filament. However, by stacking a large number of filaments, in
 the Two-degree Field Galaxy Redshift Survey (2dFGRS), it has been
 shown that the star formation in galaxies is enhanced
 3--4\,Mpc from the cores of the clusters, suggesting
 that occurrence of burst of star formation in infalling galaxies
 should be a common phenomenon \citep{p2,p08}.  

 Since outside the virial boundaries of clusters, the density of the ICM is
 too low to influence the incoming galaxies, it was suggested in these
 studies that the enhancement in star formation is due to high speed
 encounters occurring amongst infalling galaxies on the outskirts of
 galaxy clusters. 
 Multiple starburst galaxies have been seen in similar single systems associated
 with filaments, such as in Abell 1763 \citep[$z = 0.23$;][]{fadda}.
 Multi-wavelength studies of photometric properties of galaxies belonging to the Shapley
 supercluster ($z=0.048$) revealed that the supercluster environment might be
 solely responsible for the evolution of faint ($\gtrsim$ $M^{*}+2$) galaxies
 \citep{hb}. 

 This paper addresses the issue of enhancement in star formation
 activity in galaxies found in the infall regions of nearby galaxy
 clusters. The star formation activity is quantified using the
 integrated SFR and specific star formation rate (\ssf) of galaxies in
 a representative sample of local clusters, and emission line ratios
 to separate out the emission dominated by the active galactic nuclei
 (AGN) from that of star formation. A catalogue of large-scale
 inter-cluster filaments of galaxies, sourced from the SDSS, is used to
 evaluate the impact of the position of a galaxy in the cosmic web on
 its star formation properties. 

The outline of the paper is as
 follows: the data and sample properties are elaborated in the
 following section. The star formation properties of cluster galaxies
 are described in \S\ref{analysis}, briefly discussing the behaviour
 of galaxies falling in the cluster for the first time. The impact of
 the immediate and the large-scale structure (LSS) environment on the star formation
 properties of galaxies is discussed in \S\ref{discussion}, and the
 main results are summarised in \S\ref{conclusions}.  Throughout this
 work a cosmology with $\Omega_{m}=1$ \& $\Omega_{\Lambda}=0$ and
 $h=70$ km~s$^{-1}$ Mpc$^{-1}$ was adopted for calculating distances
 and absolute magnitudes.  We note that in the redshift range chosen
 for this work ($0.02\leq z \leq 0.15$), the results are insensitive
 to the choice of cosmology.
  
%%%%%%%%%%%%%%%%%%%%%%%%%%%%%%%%%%%%%%%%%%%%%%%%%%%%%%%%%%%%%%%%%%%%%%%%%%%%%%%%%%%%%%%%%%%%%%%%%%%%%%%%

\section{The Sample}
\label{data}
We used the spectroscopic galaxy data provided by the SDSS
DR4~\citep{a2}, which uses two fibre-fed double spectrographs,
covering a wavelength range of 3800-9200 \AA. The resolution
$\lambda/\Delta \lambda$ varies between $1850$ and $2200$ in different
bands. The galaxy magnitudes and the corresponding galactic extinction
values in the $r$-band were taken 
from the New York University Value added galaxy Catalogue
\citep[NYU-VAGC;][]{b5}. The $k$-corrections (to $z = 0.1$) provided
by \citet{y1} were used to correct for the effect of distance on the
magnitude of galaxies.  Only those galaxies with absolute magnitude
$M_{r} \geq -20.5$ (SDSS magnitude limit at the median redshift of our
sample $z = 0.1$) were considered.

\subsection{Cluster membership}
\label{member}

All Abell clusters \citep{a1} in the redshift range $0.02<z\leq0.15$
were considered.  In order to assign galaxies to clusters differential
redshift slices were used, i.e., galaxies,
within a projected radius, $R\!=\!6$~Mpc of the cluster
centre at the redshift of the cluster, were included
if they were found to be within a redshift slice of:

 \begin{equation}
 \Delta z=\left\{
   \begin{array}{ll}  
            3000 \hbox{ km s$^{-1}$} &  \textrm{~~~if $R\leq1$~Mpc}   \\  
	    2100 \hbox{ km s$^{-1}$} &  \textrm{~~~if $1<R\leq3$~Mpc} \\
            1500 \hbox{ km s$^{-1}$} &  \textrm{~~~if $3<R\leq6$~Mpc} \\
 \end{array}
 \right\}.
 \label{eq:selection} 
 \end{equation}

 This differential velocity slice approach helps in reducing
 interlopers without following a rigorous procedure for assigning
 cluster membership, and hence is well suited for a statistical study
 like this. Note that popular methods, such as the 3$\sigma_v$ velocity
 dispersion cut, which is often employed to define galaxy membership
 of clusters, is not our preferred choice because of our need to
 include the outskirts of clusters.

 \begin{figure}
 \centering{
 {\rotatebox{270}{\epsfig{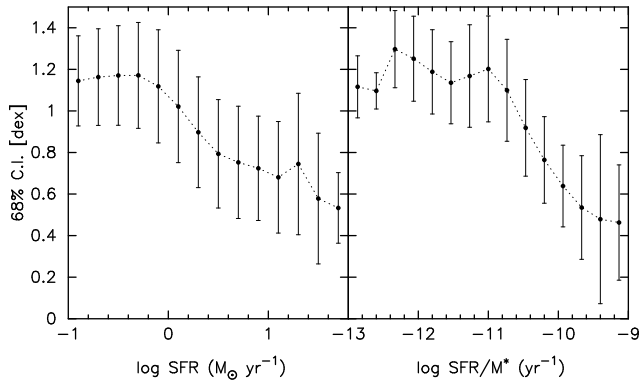}}}}
 \caption{The distribution of the errors on the SFRs of galaxies
 in our sample. The solid points denote the mean $68$\% confidence
 interval (C.I.) of the likelihood distributions for the SFRs in bins of
 log SFR ({\it{left}}) and log SFR/$M^*$  ({\it{right}}). The error
 bars show the 2$\sigma$ scatter in each bin. Galaxies with high SFR have
 higher S/N ratio in the emission lines, and hence tighter constraints
 on the estimated SFR and SFR/$M^*$ relative to the galaxies with low
 or no optical emission. It is also noticeable that the scatter in the C.I. in all
 the bins of SFR (and log SFR/$M^*$) is comparable.}
 \label{error}
 \end{figure}
 
 \subsection{Cluster parameters}
 
 Once the galaxy members of the clusters were found as described in
 \S\ref{member}, their velocity dispersion ($\sigma_v$) is determined using
 the bi-weight statistics \citep[ROSTAT,][]{b1}. Then assuming
 cluster mass $M(\rm{r})\propto \rm{r}$, r$_{200}$ (the radius at which the mean
 interior over-density in a sphere of radius r is $200$ times the critical
 density of the Universe) was calculated, 
using the relation given by \citet{c1},
 \begin{equation}
 \rm{r}_{200}=\frac{\sqrt{3}}{10}\frac{\sigma}{H(z)}.
 \end{equation}   
 Any comparison clusters (described below) which were not fully
 covered in DR~4 were excluded along with any clusters for which
 reliable velocity dispersion information was neither available nor
 obtainable from the given data.

 The final sample consists of $107$ Abell clusters \citep{a1} with
 $z<0.15$ lying in those regions of the sky where SDSS DR~4 has more
 than 70\% spectroscopic coverage.

\subsection{Galaxy data}
\label{galaxy}

The galaxy data (SFR, specific star formation rate, SFR/$M^*$) used in
this paper were sourced from the catalogue of \citet[][B04
henceforth]{b2} which is based on SDSS DR4. The Bayesian technique
employed by B04 for modelling various physical parameters including
SFR and SFR/$M^*$ is advantageous over using a single emission line
such as H$\alpha$ or [OII] as a star formation indicator because the
integrated SFR is a synthesized representation of various line
indices.  This reduces the uncertainty inherent in using a single
index by several orders of magnitude.  The simultaneous use of
different indices enables us to separate the galaxies with a
significant AGN component from those with emission entirely due to
star formation (see below). Using the entire spectrum also makes it
possible to obtain an independent estimate of the stellar mass $M^*$
(from the $z$-band photometry), which further constraints the
estimated SFR/$M^*$, making it a more reliable parameter for
evaluating star formation properties of galaxies, relative to
comparable luminosity based parameter used elsewhere
\citep[\eg][]{lewis}.

B04 divided all galaxies into star forming, AGN and composite using
the BPT \citep{b3} diagram and then derived the SFR and SFR/$M^*$ from
the spectra taken by the (3$^{\prime\prime}$ diameter) SDSS
fibre. These SFRs and SFR/$M^*$ were then corrected for aperture
biases using galaxy colours measured `globally', and within the fibre
(see B04 for details).  Out of the three statistical estimates (mean,
median and mode) for the probability distribution of SFR and SFR/$M^*$
derived for each galaxy, we used the median of the probability
distribution function (PDF) of the star formation parameters since it
is independent of binning.

\begin{figure*}
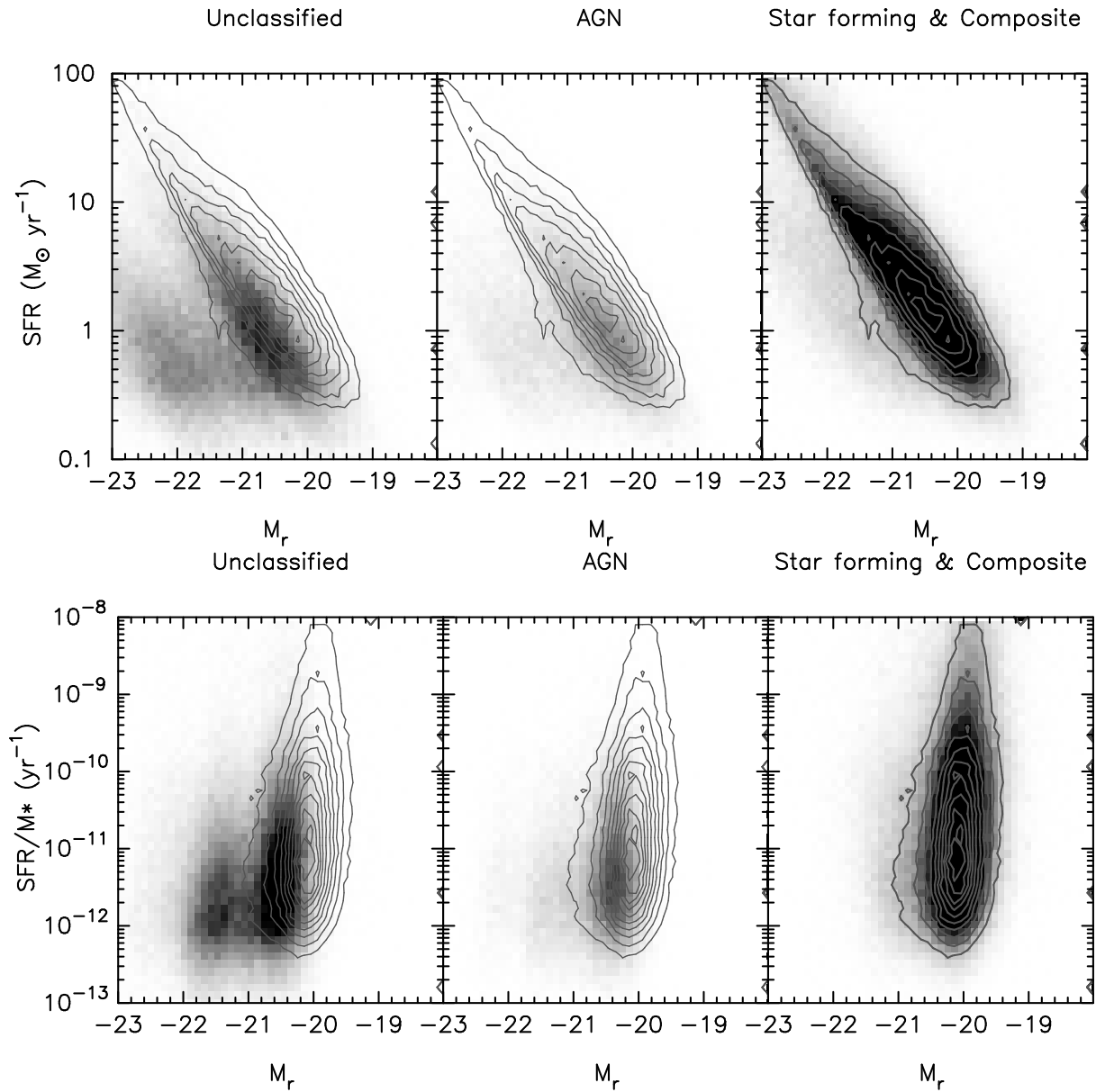

\centering{
{\rotatebox{270}{\epsfig{file=figure2a.ps,width=8cm}}}}
\vfill
\centering{
{\rotatebox{270}{\epsfig{file=figure2b.ps,width=8cm}}}}
\caption{{\it{(top)}} SFR and {\it{(bottom)}} SFR/$M^*$ as a function
  of $r$-band magnitude for all the galaxies in SDSS DR4, divided
  according to their position in the BPT diagram (see
  \S\ref{galaxy}). The contours for star-forming+Composite class
  (right panel) are repeated in other panels to compare the relative
  distributions. The outermost contour in each case represents 100
  galaxies per bin, increasing by 100 for every consecutive contour
  inwards. As expected, the AGN hosts follow the passive galaxy
  distribution. It is also interesting that the difference between
  various classes become more distinct for SFR/$M^*$ (bottom panel).
  These plots show that absolute SFR follows the galaxy luminosity,
  but the trend reverses when SFR/$M^*$ is considered. Another
  interesting feature that emerges here is that the observed
  bimodality in SFR seems to largely consist of the
  `non-emission line' galaxies. This plot gives no indication of
  enhanced (or suppressed) star formation in AGN hosts, although this might be a result of
  the obscuration of emission lines by dust around the active nuclei.}
\label{mag-sf}
\end{figure*}
  
 Fig.~\ref{error} shows the 68\% confidence intervals (C.I.) for log SFR
 estimate for each galaxy in our sample and the scatter in C.I. in bins of log
 SFR and log SFR/$M^*$ respectively. The trends are similar to those seen in
 fig.~14 of B04 (for all galaxies in SDSS DR~4). Since the SFR for the 
 star-forming galaxies are measured directly from the high S/N emission
 lines ($>3$ in all 4 lines: H$\alpha$, H$\beta$, [OIII] and [NII] required for
 the BPT diagram), the constraint on the derived value of SFR is tight.
 However, for the low S/N and non-emission line galaxies, most of which are
 passively evolving, SFR is measured indirectly from the $4000$\,\AA~break-SFR
 relation obtained for the high S/N galaxies (see B04 for details), resulting in high
 uncertainties in the estimated SFR and \ssf. We chose not to
 select galaxies based on the confidence interval values derived from the
 probability distribution function of the SFR and \ssf~parameters of
 individual galaxies because excluding galaxies with higher uncertainties
 in \ssf~(or SFR) distribution would exclude a significant number of
 passive galaxies from the sample. But precautionary measures are taken into account
 where necessary (\S\ref{unclass}), so that our results remain virtually
 unaffected by the uncertainties in the \ssf~(or SFR) estimates for the
 unclassified galaxies.

 \subsection{Unclassified galaxies}
 \label{unclass}

 Any galaxy which can not be classified using the BPT diagram,
 because of low S/N ratio or non-detection of the requisite emission lines, 
 is termed `unclassified' in B04, which makes this class of galaxies
 rather heterogeneous.  
The spectra obtained from the 3$^{\prime\prime}$ fibre, which encompasses
 $\sim$ 40-60\% of the galaxy's light (at $z\sim0.1$), may hence only represent
 the passive core of the galaxy, lacking any emission lines. 
 A careful analysis of the images of these galaxies revealed that these are predominantly
bulge dominated systems, in fact even more concentrated
 than AGN or composite galaxies (J.~Brinchmann, private communication).
  A detailed investigation of whether all of these galaxies are indeed
  passively evolving or comprise a sub-population of dust obscured
  galaxies with an AGN supported star formation in the galactic
  nucleus is beyond the scope of this work \citep[but see][where we
  briefly address related issues]{mahajan09}.

 Fig.~\ref{mag-sf} shows the distribution of absolute SFR
 and log SFR/$M^*$ of all galaxies ($-23\leq M_{r}\leq-18$) in SDSS DR4
 as a function of their luminosity for the three different classes:
 star forming (\& composite), AGN and the unclassified (B04). As
 previously speculated, the unclassified galaxy class seems
 to be dominated by passive galaxies. It is also not
 surprising that the galaxies hosting an AGN which are known to be
 bulge-dominated systems, follow the same distribution as that of
 the unclassified galaxies.

 In agreement with previous findings, Fig.~\ref{mag-sf} shows that
 most low-luminosity galaxies are efficiently forming stars
 \citep{k3,h2,h1,ssf,mahajan10}. Interestingly, the observed
 bi-modality in the distribution of SFR is exclusive to the
 unclassified galaxies.
 If the SFR of the star-forming galaxies is a function of environment,
 one should expect the distribution of their SFR to show a bimodal
 distribution in a dataset which samples a wide range of environments,
 such as that employed here. However, no conclusive remarks can be
 made on the issue of starburst-AGN connection based on the analysis
 presented here, because at visible wavelengths, the circum-nuclear star
 formation (which largely contributes to the 3$^{\prime\prime}$ SDSS
 fibre spectra) may be entirely obscured by dust enshrouding the
 active nucleus \citep[\eg][]{popescu05,prescott07}.

\subsection{Starburst galaxies}
\label{starburst}

\begin{figure}
 \centering{
{\rotatebox{270}{\epsfig{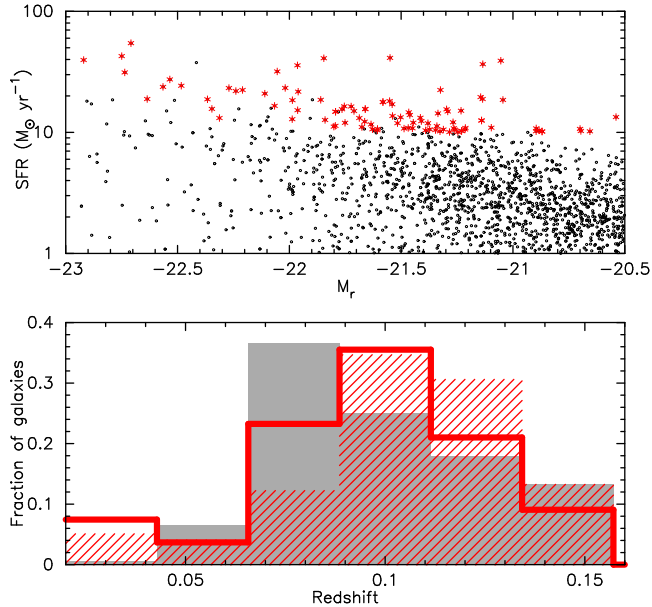}}}}
\caption{{\it{Top:}} Distribution of all {\it{(black points)}} cluster galaxies and starbursts {\it{(red stars)}}
within 3r$_{200}$ of the cluster centres in the SFR-$M_r$ space shows that the starburst
galaxies span the entire magnitude range as spanned by other cluster galaxies.
 {\it{Bottom:}} The redshift distribution of galaxies in the CN {\it{(solid grey)}} and SB {\it{(open red)}}
 cluster samples. The K-S test does not distinguish between the parent population of these two samples.
 The {\it{red hatched}} distribution is for the starburst galaxies in the SB sample.}
\label{z-mag}
\end{figure}

 A galaxy having current SFR much higher than that averaged over its past lifetime may be regarded
 as a \s. In the context of this paper, dual criteria were adopted to define a \s~galaxy:
 log SFR/$M^*\geq-10$ yr$^{-1}$ {\it{and}} SFR $\geq10$ M$_{\odot}$yr$^{-1}$. These thresholds
 are purely empirical, driven by the idea of selecting galaxies having current SFR
 much higher than that averaged over their past star formation history. These limits
 thus select galaxies contributing to the high end tail of the SFR and SFR/$M^*$
 distributions only, as shown in Figure~\ref{z-mag} (top panel).
  
 In particular, they label $\sim 3\%$ of all and $\sim 13\%$ of star-forming galaxies in
 our sample of cluster galaxies as a \s. In agreement with several other studies
 \citep[\eg][]{feulner05,noeske07}, we note that by choosing \s~galaxies according
 to absolute SFR or SFR/$M^*$ alone could bias the selected galaxies to giant or dwarf
 galaxies respectively \citep[because dwarf galaxies have higher SFR/$M^*$ than giants
 but lower absolute SFR; \eg][]{ssf}. These \s~galaxies also cover a wide
 range in structural properties. %; some typical examples found on the cluster outskirts
% (1-2r$_{200}$) are shown in Fig.~\ref{images}. 
 
 We stress that the above definition of \s~galaxies is used only to sub-classify galaxy
 clusters in our sample. The results presented in this paper remain unaffected by the
 specific choice of SFR and \ssf~thresholds quantitatively (see \S\ref{analysis} for a 
 detailed discussion).

   \begin{figure}
 \centering{
{\rotatebox{270}{\epsfig{file=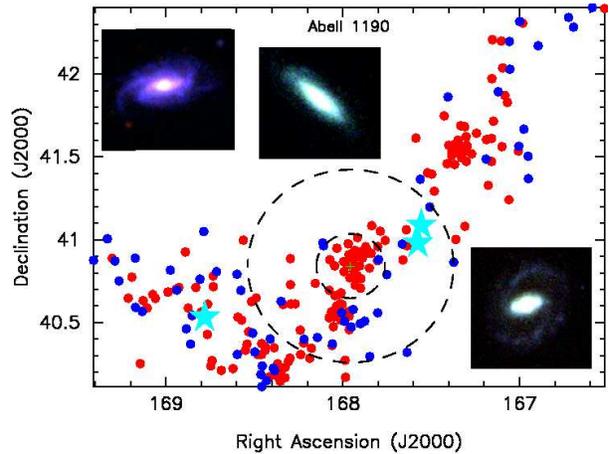,width=6.0cm}}}}
\caption{The distribution of {\it{red}} and {\it{blue}} galaxies (colour-selected in 
 the colour-magnitude space; see text) in and
 around the rich cluster Abell~1190, which is part of our SB sample.
 The starburst galaxies in its vicinity are shown as cyan {\it{stars}} and
 the corresponding optical images are shown inset. The dotted circles
 represent regions encompassing 1 and 3 Mpc from the centre of the
 cluster ({\it{green cross}}). }
\label{sb}
\end{figure} 
 
%%%%%%%%%%%%%%%%%%%%%%%%%%%%%%%%%%%%%%%%%%%%%%%%%%%%%%%%%%%%%%%%%%%%%%%%%%%%%%%%%%%%%%%%%%%%%%%%

\section{Star formation and Environment}
\label{analysis}

\begin{table}
\caption{Abell clusters with starburst galaxies on outskirts (SB)}
\label{sb-table}
\begin{minipage}{\linewidth}
\begin{tabular}{|c|r|r|c|c|c|c|}
\hline
 Abell  & RA      & Dec    & Richness & z  & Velocity     \\
       & (J2000) & (J2000)&          &    & dispersion    \\
       &         &        &          &    & ($\sigma_v$; km s$^{-1}$) \\ \hline
  152  &   1  9 50.0  & +13 58 45 &   0 &  0.0581 & 756  \\
  175  &   1 19 33.0 &  +14 52 29 &   2 &  0.1288 & 705  \\
  628  &   8 10  9.5 &  +35 12 28 &   0 &  0.0838 & 741  \\
  646  &   8 22 12.0 &  +47  5 44 &   0 &  0.1303 & 849  \\
  655  &   8 25 23.1 &  +47  6 32 &   3 &  0.1267 & 858  \\
  714  &   8 54 50.2 &  +41 54  0 &   1 &  0.1397 & 609  \\
  724  &   8 58 20.4 &  +38 33 48 &   1 &  0.0934 & 492  \\
  856  &   9 45 30.1 &  +56 32 43 &   0 &  0.1393 & 450\footnote{\citet{p1}}  \\
  874  &   9 50 43.8 &  +58  1 29 &   1 &  0.1480 & 1551 \\
  941  &  10  9 42.7 &  + 3 40 56 &   1 &  0.1049 & 195  \\
  971  &  10 19 49.6 &  +40 57 34 &   1 &  0.0929 & 873  \\
  975  &  10 22 47.3 &  +64 37 29 &   2 &  0.1182 & 273  \\
 1004  &  10 25 36.5 &  +51  3 23 &   1 &  0.1414 & 963  \\
 1021  &  10 28 45.2 &  +37 39 21 &   1 &  0.1114 & 828  \\
 1076  &  10 45  9.5 &  +58  9 55 &   1 &  0.1164 & 507  \\
 1132  &  10 58 25.5 &  +56 47 41 &   1 &  0.1363 & 792  \\
 1190  &  11 11 49.2 &  +40 50 30 &   2 &  0.0752 & 693  \\ 
 1227  &  11 21 38.4 &  +48  2 22 &   2 &  0.1120 & 813  \\
 1238  &  11 22 58.6 &  + 1  6 21 &   1 &  0.0733 & 552  \\
 1341  &  11 40 35.8&   +10 23 18 &   1 &  0.1050 & 834  \\
 1342  &  11 40 41.8 &  +10  4 18 &   1 &  0.1061 & 390  \\
 1346  &  11 41 10.9 &  + 5 41 18 &   1 &  0.0976 & 711  \\
 1354  &  11 42 11.7 &  +10  9 14 &   1 &  0.1178 & 456  \\
 1373  &  11 45 27.8 &  - 2 24 40 &   2 &  0.1314 & 627  \\
 1377  &  11 47  2.4 &  +55 44 16 &   1 &  0.0511 & 684  \\
 1424  &  11 57 34.1 &  + 5  2 14 &   1 &  0.0774 & 699  \\
 1446  &  12  1 55.9 &  +58  1 14 &   2 &  0.1031 & 747  \\
 1496  &  12 13 26.2 &  +59 16 19 &   1 &  0.0937 & 396  \\
 1518  &  12 19  3.9&   +63 30 24 &   0 &  0.1080 & 783  \\
 1534  &  12 24  7.5 &  +61 30 26 &   0 &  0.0698 & 321  \\
 1544  &  12 27 46.3 &  +63 25 25 &   1 &  0.1454 & 597  \\
 1713  &  13 19 11.8 &  +58  5 22 &   1 &  0.1406 & 1095 \\ 
 1738  &  13 25  9.7 &  +57 36 35 &   2 &  0.1155 & 597  \\
 1750  &  13 30 52.4 &  - 1 51  6 &   0 &  0.0855 & 795  \\
 1896  &  14 18 40.9 &  +37 48 27 &   0 &  0.1329 & 534  \\
 1999  &  14 54  5.6 &  +54 20  5 &   1 &  0.0994 & 333  \\
 2026  &  15  8 33.8 &  - 0 15 51 &   1 &  0.0875 & 753\footnote{\citet{p1}}  \\
 2051  &  15 16 46.4 &  - 0 56 28 &   2 &  0.1155 & 600  \\
 2053  &  15 17 15.8 &  - 0 40 23 &   1 &  0.1127 & 831  \\
 2142  &  15 58 16.3 &  +27 13 53 &   2 &  0.0894 & 942  \\
 2149  &  16  1 37.9 &  +53 52 57 &   0 &  0.1068 & 723  \\
 2183  &  16 21 31.6 &  +42 43 20 &   1 &  0.1366 & 495  \\
 2197  &  16 28 10.3 &  +40 54 47 &   1 &  0.0301 & 564  \\
 2199  &  16 28 36.8 &  +39 31 48 &   2 &  0.0299 & 759  \\
 2244  &  17  2 44.2 &  +34  4 13 &   2 &  0.0970 & 1164 \\
 2255  &  17 12 30.5 &  +64  5 33 &   2 &  0.0809 & 1116 \\ \hline 
\end{tabular}
\end{minipage}
\end{table} 

\begin{table}
\caption{Comparison (CN) clusters}
 \label{cn-table}
\begin{minipage}{\linewidth}
\begin{tabular}{|c|r|r|c|c|c|c|}
\hline
Abell  & RA          & Dec    & Richness & z  & Velocity        \\
       & (J2000)     & (J2000)&          &    & dispersion      \\
       &             &        &          &    & ($\sigma_v$; km s$^{-1}$)   \\ \hline
 190    &  1 23 41.9  & - 9 51 30 &   0   &  0.1015 &         354\\
 612    &  8  1  4.0  & +34 48  2 &   1   &  0.0774 &         186\\
 667    &  8 28  6.0  & +44 42 23 &   0   &  0.1450 &         792\\
 671    &  8 28 31.0  & +30 24 26 &   0   &  0.0503 &         801\\
 690    &  8 39 16.0  & +28 49 50 &   1   &  0.0788 &         516\\
 695    &  8 41 26.1  & +32 16 40 &   1   &  0.0687 &         375\\
 775    &  9 16 21.2  & + 5 52  2 &   1   &  0.1340 &         975\\
 819    &  9 32 17.9  & + 9 39 18 &   0   &  0.0764 &         561\\
 858    &  9 43 26.6  & + 5 52 52 &   0   &  0.0881 &         918\\
 990    & 10 23 33.8  & +49  9 30 &   1   &  0.1440 &        1041\\
 1024   & 10 28 18.5  & + 3 45 21 &   1   &  0.0743 &         690\\
 1038   & 10 32 59.4  & + 2 15 17 &   1   &  0.1246 &         297\\
 1064   & 10 38 46.9  & + 1 16  4 &   1   &  0.1320 &         594\\
 1068   & 10 40 50.2  & +39 57  1 &   1   &  0.1386 &        1317\\
 1080   & 10 43 58.8  & + 1  5  0 &   0   &  0.1205 &         432\\
 1139   & 10 58  4.9  & + 1 30 45 &   0   &  0.0395 &         249\\
 1169   & 11  8 10.0  & +43 56 34 &   1   &  0.0587 &         591\\
 1171   & 11  7 29.2  & + 2 56 31 &   0   &  0.0757 &         159\footnote{\citet{p1}}\\
 1173   & 11  9 14.4  & +41 34 33 &   1   &  0.0759 &         570\\
 1205   & 11 13 23.0  & + 2 30 29 &   1   &  0.0759 &         810\\
 1225   & 11 21 24.5  & +53 46 22 &   0   &  0.1040 &         732\\
 1270   & 11 29 33.3  & +54  4 20 &   0   &  0.0686 &         600\\
 1302   & 11 33 30.4  & +66 25 18 &   2   &  0.1160 &         834\\
 1318   & 11 36 30.6  & +54 58 16 &   1   &  0.0564 &         366\\
 1361   & 11 43 48.3  & +46 21 17 &   1   &  0.1167 &         453\\
 1364   & 11 43 39.8  & - 1 46 36 &   1   &  0.1063 &         558\\
 1368   & 11 45  1.7  & +51 15 13 &   1   &  0.1291 &         762\\
 1372   & 11 45 29.3  & +11 31 13 &   1   &  0.1126 &         435\\
 1383   & 11 48 13.5  & +54 37 12 &   1   &  0.0600 &         426\\
 1385   & 11 48  5.5  & +11 33 16 &   1   &  0.0831 &         531\\
 1387   & 11 48 54.1  & +51 37 12 &   1   &  0.1320 &         753\\
 1390   & 11 49 35.1  & +12 15 15 &   0   &  0.0829 &         342\\
 1402   & 11 52 37.4  & +60 25 15 &   0   &  0.1058 &         219\\
 1452   & 12  3 42.6  & +51 44 14 &   0   &  0.0628 &         513\footnote{\citet{struble}}\\
 1552   & 12 29 50.6  & +11 44 29 &   1   &  0.0859 &         711\\
 1559   & 12 33 10.0  & +67  7 31 &   1   &  0.1066 &         693\\
 1564   & 12 34 57.3  & + 1 51 32 &   0   &  0.0792 &         687\\
 1566   & 12 35  5.1  & +64 23 32 &   2   &  0.1000 &         561\\
 1577   & 12 37 52.0  & - 0 16 22 &   1   &  0.1405 &         366\\
 1621   & 12 48 45.9  & +62 41 46 &   1   &  0.1029 &         534\\
 1630   & 12 51 44.8  & + 4 34 49 &   1   &  0.0648 &         426\\
 1637   & 12 53 59.3  & +50 49 51 &   1   &  0.1270 &         564\\
 1646   & 12 55 48.6  & +62  9 53 &   0   &  0.1063 &         468\\
 1650   & 12 58 46.1  & - 1 45 57 &   2   &  0.0845 &         513\\
 1663   & 13  2 46.4  & - 2 31 49 &   1   &  0.0847 &         891\\
 1764   & 13 34 43.2  & +59 55 51 &   0   &  0.1167 &         585\\
 1773   & 13 42  8.5  & + 2 15  7 &   1   &  0.0766 &        1098\\ 
 1780   & 13 44 38.1  & + 2 53 12 &   1   &  0.0786 &         426\\
 1882   & 14 14 39.8  & - 0 20 33 &   3   &  0.1367 &         612\\
 1885   & 14 13 47.6  & +43 40 15 &   1   &  0.0890 &        1038\\ 
 1918   & 14 25  9.2  & +63  9 41 &   3   &  0.1394 &        1008\\
 1920   & 14 27 17.5  & +55 46 46 &   2   &  0.1310 &         570\\
 1936   & 14 34 28.8  & +54 50  5 &   1   &  0.1386 &         498\\
 2018   & 15  1 12.8  & +47 16 26 &   1   &  0.0872 &        1029\\ 
 2046   & 15 12 41.4  & +34 51  9 &   1   &  0.1489 &         948\\
 2062   & 15 21 19.8  & +32  5 38 &   1   &  0.1155 &         537\\
 2110   & 15 39 43.5  & +30 42 45 &   1   &  0.0980 &         654\\
 2175   & 16 20 23.1  & +29 55 18 &   1   &  0.0972 &         876\\
 2196   & 16 27 21.6  & +41 29 43 &   0   &  0.1332 &         687\\
 2211   & 16 34  4.0  & +40 57  9 &   1   &  0.1355 &         537\\
 2245   & 17  2 45.0  & +33 33 13 &   1   &  0.0852 &         984\\ \hline
\end{tabular}
\end{minipage}
\end{table}

\begin{figure}
  \centering{
    {\rotatebox{270}{\epsfig{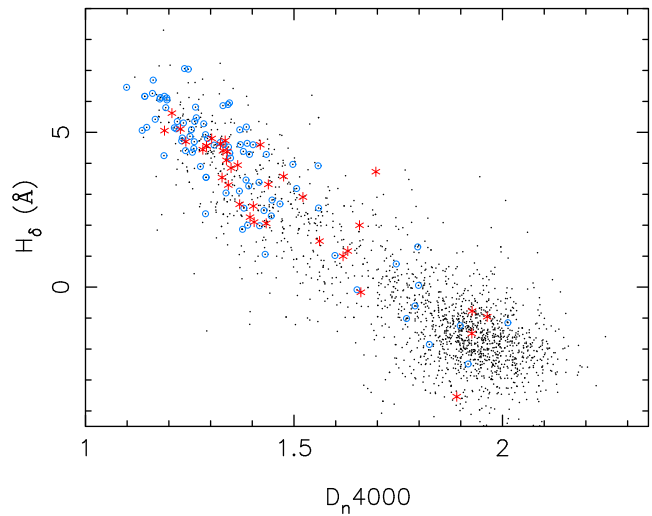}}}}
\caption{All galaxies in the SB clusters are plotted as {\it{black}} dots
 in the D$_n$4000-H$\delta$ space. Overplotted as {\it{red}} stars are starburst
 galaxies found at $1 \leq$r/r$_{200} \leq 2$. The {\it{blue}} circles are galaxies
 at the same distance from the cluster, but having $1 \leq \rm {SFR} < 10$ M$\odot$ yr$^{-1}$.
 Most of these infalling galaxies with high SFR (and SFR/$M^*$) occupy the area
 in the D$_n$4000-H$\delta$ space where galaxies with currently ongoing or
 recent starbursts are expected to lie \citep{k3}.}
\label{d4000-h}
\end{figure}

From a sample of $107$ galaxy clusters, those having at least one
($M_{r}\leq-20.5$)
starburst galaxy (as defined in \S\ref{starburst}),
{\it{anywhere}} within $3$ Mpc of the cluster centre, were
selected, yielding $46$ clusters termed as `SB clusters'.
The distance range of 3~Mpc was chosen since,
for most of the galaxy clusters in our sample, r$_{200}\sim 1.5$
Mpc. Hence the \s~galaxies chosen to define this sub-class of clusters
would either belong to the cluster or to its outskirts. 
In order to make the selection procedure independent of the
uncertainties in the SFR estimate (\S\ref{galaxy}; Fig.~\ref{error}),
\s~galaxies which are unclassified (\S\ref{unclass}) but have
estimated SFR $\geq10$ $M_{\odot}$ yr$^{-1}$ in the B04 catalogue, were
ignored. Redshift distribution of cluster galaxies in the two samples
 are shown in Figure~\ref{z-mag} (bottom panel). The Kolmogorov-Smirnov (KS)
 statistic suggests that the two samples are statistically similar. The distribution
 of redshifts for the starburst galaxies is also statistically similar to that of
 other cluster galaxies.  
 
A typical SB cluster, Abell~1190, is shown in Fig.~\ref{sb} as an
example. In this figure, galaxies are labelled as red or blue using a
colour-magnitude relation fitted to galaxies within $3$\,Mpc. Of the
rest of the 61 clusters which do not have any starburst galaxy within
3 Mpc of the cluster centre form the `comparison' (CN) cluster
sample. The fundamental properties of our cluster samples are listed
in Tables \ref{sb-table} and \ref{cn-table} 
respectively. In total 3,966 galaxies are found within 3r$_{200}$ of
the cluster centres, of which 3,903 are non-AGN and were used for all
further analysis, unless stated otherwise.
If the occurrence of \s~galaxies in some clusters is purely
coincidental, or the result of line-of-sight superposition, then CN
and SB cluster samples should display similar properties. As shown
below, this hypothesis fails to hold.
 
 \subsection{The last `bursts' of star formation}

 The $4000$\,\AA~break (D$_{n}$4000) is a prominent discontinuity
 occurring in the continuum of the visible spectrum of a galaxy.  It
 arises because a large number of absorption lines, mainly from
 ionised metals accumulate in a small wavelength range. In hot young
 stars, the metals are multiply ionised and hence the opacity is low,
 resulting in a smaller D$_{n}$4000 relative to old, metal rich
 stars. We used the same definition of D$_{n}$4000 as adopted by B04,
 which is the ratio of average flux density in the narrow continuum
 bands (3850-3950 and 4000-4100\,\AA). This definition was initially
 introduced by \citet{balogh99} and is an improvement over the
 original definition of \citet{bruzual83}, in the sense that the narrow
 continuum bands make the index considerably less sensitive to reddening
 effects.

 The H$\delta$ absorption line occurs when the visible light of a
 galaxy mainly comes from late-B and early-F stars. A large absorption
 in H$\delta$ implies that a galaxy experienced a major burst of star
 formation $\sim$\,0.1--1 Gyr ago, while a stronger D$_n$4000 is an
 indicator of metal-rich ISM. Hence together EW(H$\delta$) and
 D$_n$4000 indicate whether a galaxy has been forming stars
 continuously (intermediate/high D$_n$4000 and low EW(H$\delta$)) or
 in bursts (low D$_n$4000 and high EW(H$\delta$)) over the past Gyr
 \citep{k4}. Fig.~\ref{d4000-h} shows the distribution of spectacular
 starburst galaxies (SFR$\geq10$ M$_\odot$ yr$^{-1}$ \& log
 SFR/$M^*\geq-10.5$ yr$^{-1}$), relative to all other galaxies in the
 SB cluster sample, in the D$_{n}4000$-H$\delta$ plane. It is clear
 that our adopted criteria for defining starburst galaxies on the
 basis of integrated SFR and SFR/$M^*$ segregates well these galaxies
 in the space defined by spectral indices sensitive to recent star
 formation. For a comparison, we also show galaxies which have the
 same SFR/$M^*$ but absolute SFR which is upto 10 times lower than
 `starburst' galaxies. For the latter, the distribution shows a larger
 scatter.
 
 \subsection{Starburst galaxies on the outskirts of clusters}
 \label{sbg}
 
 \begin{figure*}
\centering{
{\rotatebox{270}{\epsfig{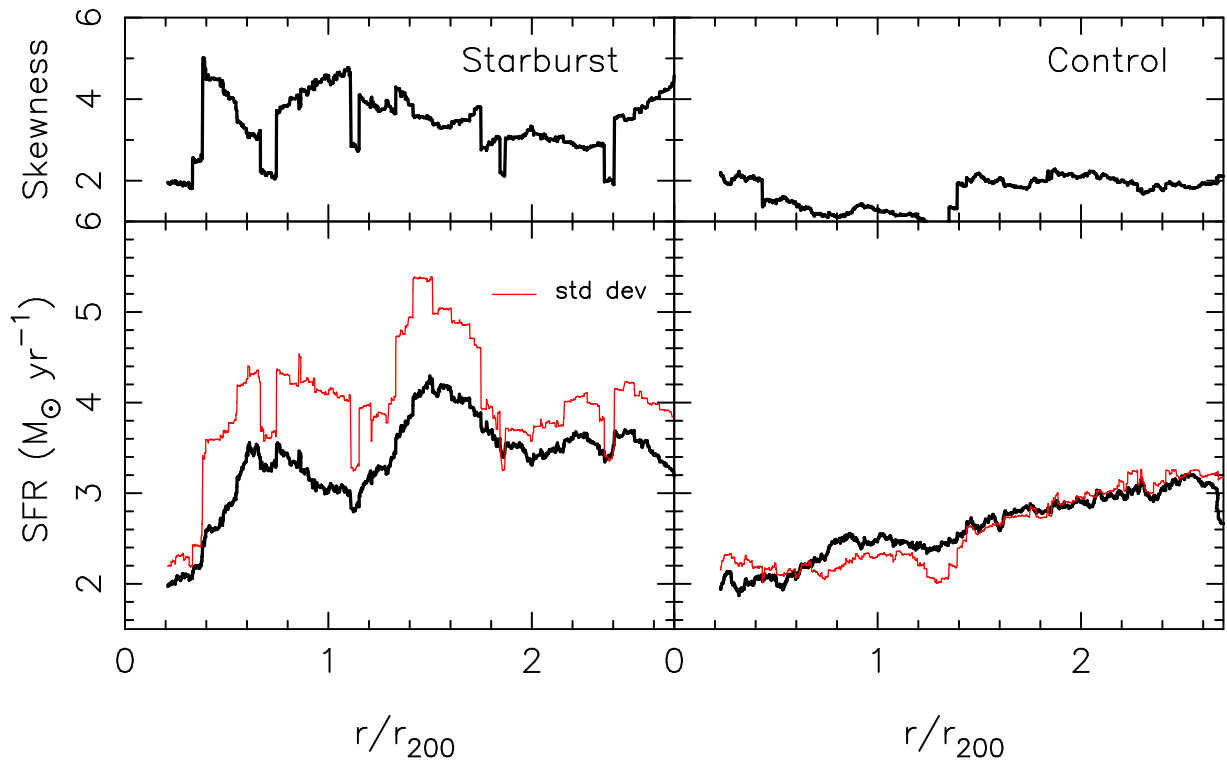}}}}
\caption{The mean ({\it{thick line, bottom panel}}), skewness
 ({\it{upper panel}}) and standard deviation ({\it{thin red, bottom panel}}) in SFR of
 galaxies in SB ({\it{left}}) and CN ({\it{right}}) clusters in running
 bins ($250$ galaxies per bin) of (scaled) cluster-centric radius.
 Unclassified galaxies with SFR$> 10$ M$_{\odot}$ yr$^{-1}$ and AGN were excluded.
 As expected, the SFR falls toward the cluster centre. But the difference
 in the SFR profiles of the two samples is striking. The high standard
 deviation at the cluster boundary (r $\sim$ r$_{200}$) indicates the presence of
 galaxies with extreme SFRs. It is also noticeable
 that the mean SFR of CN clusters is lower than that of SB clusters in general.}
\label{sfr}
\end{figure*}

We began this section by sub-classifying clusters on the basis of the
presence of starburst galaxies in them. In order to test the
hypothesis that these \s~galaxies might occur randomly within a
cluster, we plot the first, square root of the second and the third
moment (mean, standard deviation \& skewness) of SFR respectively, as
a function of (scaled) cluster-centric radius, in Fig.~\ref{sfr}. We
extend our analysis upto $3\,r_{200}$ because it encompasses the
core of the galaxy cluster and its outskirts \citep{dg,lewis,g3},
providing a range of galactic environments to explore.
 
The SFR profiles of the two samples in Fig.~\ref{sfr} 
are strikingly different, with SB
clusters showing a higher mean SFR relative to CN clusters at almost
all radius.  The striking feature of this stacked SFR profile is the
sudden increase in the standard deviation in SFR, which, accompanied
by positive skewness, indicates an enhancement in the mean star
formation of galaxies in the SB clusters.  This shows that several
highly star-forming galaxies are present just outside the boundary
($\sim$1.5r$_{200}$) of SB clusters. Given that the selection criteria
demanded that \s~galaxies could occur in these clusters anywhere
within the given threshold (3 Mpc), it seems highly unlikely that
occurrence of such a `blip' is a random chance.  There seems to be
another region of increase in the mean SFR of galaxies in the SB
clusters at $\sim 0.5$ r$_{200}$. We note that the latter feature,
closer to the cluster core, is
neither due to the presence of group-like structures ($\sigma_v \leq
300$ km s$^{-1}$; table~\ref{sb-table}), nor to the nearby
interacting cluster pair Abell 2197 and 2199 present in our
sample. Both these features in the SFR profile of SB clusters are thus
considered real. We discuss them in detail in the following sections.

 \subsection{Clusters \& Filaments}
 
\begin{figure*}
\centering{
{\rotatebox{270}{\epsfig{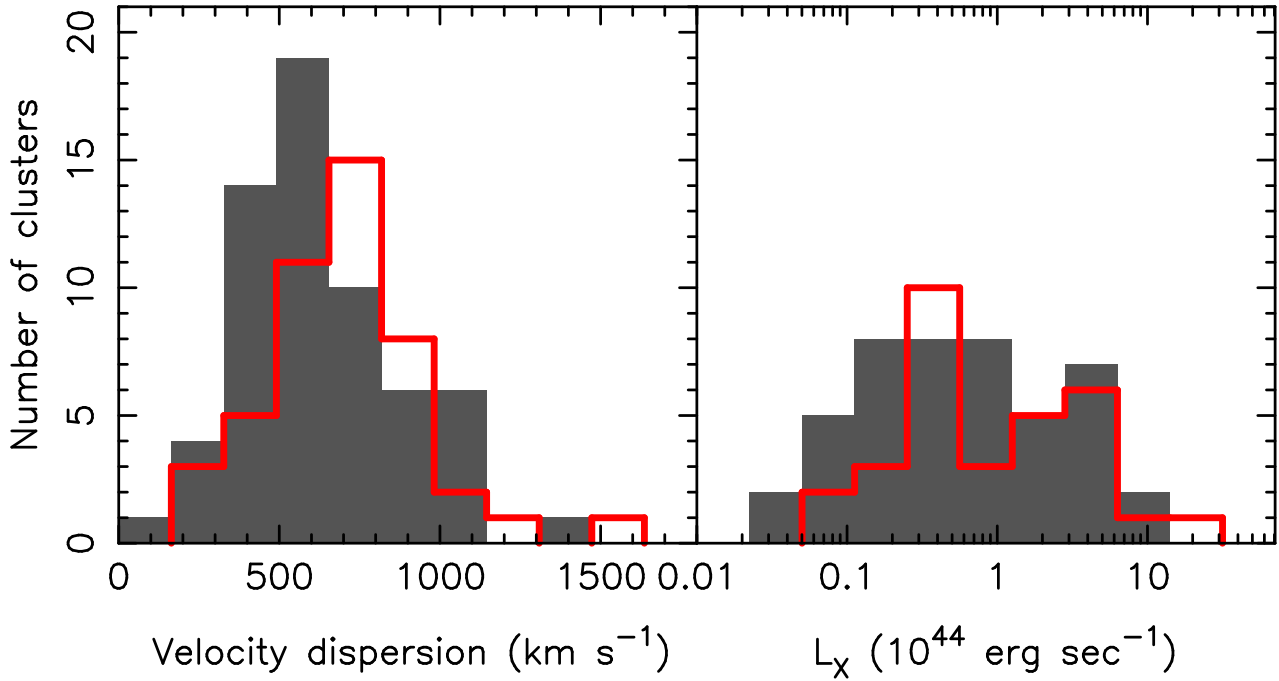}}}}
\caption{ The velocity dispersion {\it{(left)}} and  X-ray luminosity ($L_X$) {\it{(right)}} 
 distribution of SB ({\it{red open}}) and CN ({\it{grey solid}}) cluster samples respectively.
 The $L_X$ are available for almost 70\% of both samples.
While the SB clusters are found to have systematically higher velocity dispersions than
 the CN clusters, the distribution of $L_X$ show no significant statistical differences, suggesting
 that the SB clusters although not more massive than their CN counterparts are dynamically
 more unrelaxed. }
\label{veld-dist}
\end{figure*}

 In order to explore whether the enhancement in SFR of galaxies found
 on cluster outskirts (\S\ref{sbg}) is related to the properties of the
 clusters themselves, the distribution of velocity dispersion ($\sigma_v$)
 and X-ray luminosities (\lx), where available, for the two cluster samples are plotted in
 Fig.~\ref{veld-dist}. Since $\sigma_v$ are derived from spectroscopic
 redshifts using an iterative procedure \citep[\textsc{ROSTAT};][]{b1}, the error in
 individual $\sigma_v$ values is very small.  \lx~are obtained for 31/46 SB and 45/61 CN
 clusters respectively from BAX (http://bax.ast.obs-mip.fr/), which is an online database
 providing observational X-ray measurements for clusters of galaxies observed by
 various space and ground-based X-ray missions.  In the following we use \lx~measured in the
 1.5--2.4\,keV energy band as a proxy for cluster mass, and $\sigma_v$ as a proxy for
  the dynamical state of the cluster. 
  
 Fig.~\ref{veld-dist} shows that despite the fact that 
 cluster samples consist of a wide range of objects (group-like structures to massive
 clusters), their $\sigma_v$ distributions are significantly
 different. As suggested by the KS statistic, SB
 clusters have higher $\sigma_v$ than CN clusters, but there are no statistical
 differences in the distributions of the \lx~for the two samples. 
 To confirm that we are not selecting only the X-ray bright clusters in either of the samples, 
 we further checked that the $\sigma_v$ distributions of clusters from the two 
 samples, for which \lx~are not available, are not statistical different, as
 confirmed by the KS test. This implies that there is no difference in the mass distribution of
 the two samples. However, similar  \lx~but higher $ \sigma_v$ suggests that the SB clusters
 are likely to be dynamically unrelaxed relative to the CN sample. The observed
 difference in the  $\sigma_v$ distribution is not due to the two sample of clusters
 being sampled from different ranges of redshift since both the
 cluster samples cover a redshift range (0.02--0.15) spanning only $\sim$\,1.5 Gyr in
 lookback time.

If a cluster is being assembled via galaxies falling in through large-scale
 filaments, the unrelaxed dynamical state of the cluster will be observable as
 high $\sigma_v$, especially if one or more such filaments align along the line
 of sight. Hence the above observations could be interpreted as the manifestation of
 systematic differences in the dynamical state of the two cluster samples with similar
 mass distributions. Such an impact of the global environment, in which a galaxy is embedded, on
its star formation properties, is inevitable
 \citep[\eg][]{hb,p2,fadda,p08,edwards10}. However, an interesting
question is whether the large-scale structure directly influences the evolution
 of galaxies therein, or
has an indirect effect. Here we use the position of galaxies along the
cosmic-web to explore this further. The catalogue of galaxy filaments
used in this paper is sourced from SDSS DR4 by using the algorithm
described by \citet[][PDH04 thereafter]{pdh}. 

Just as in PDH04, all
filaments are classified into morphological classes from types I to V
i.e. straight, warped, sheet-like, uniform and
irregular distribution of galaxies respectively, with respect to an
inter-cluster axis. The algorithm is well described in PDH04, but for
the benefit of reader we briefly describe it here: to identify galaxy
filaments in a large-scale survey such as the SDSS involved setting up
orthogonal planar pair projections of the filaments viewed along the
inter-cluster axis. The planes corresponded to $\pm 20$ Mpc from the
inter-cluster axis and had a depth of $\pm 5$ Mpc around it. 
 This choice of depth and width ensures that the
ability to visualise highly curved and complex filaments is
retained (fig.~1 of PDH04 illustrated this process in more
detail).  Finally, those filaments for which both Mahajan and
 Pimbblet agreed on the same morphological class were included in the
 `clean' sample, and were used in all further analysis.
   
\begin{figure}
\centering{
{\epsfig{file=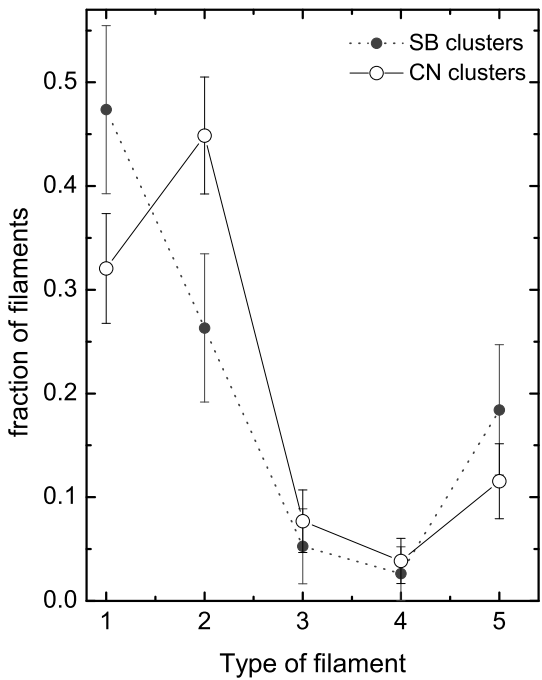,width=8.5cm}}}
\caption{The distribution of the morphological type of filament (according
  to \citet{pdh} classification) for the `clean' filaments in the SB
  ({\it{dotted}}) and CN ({\it{solid}}) cluster samples
  respectively. The errors are calculated using
  Eqn.\ref{frac-err}. The SB and CN clusters show significantly
  different preferences for type I and type II filaments
  respectively.}
\label{type}
\end{figure}

 Simulations have shown that properties, such as matter density of
 filaments, are closely related to their morphology \citep{col05,
   kd}. Therefore, by studying the morphology of filaments through
 which a galaxy approaches the cluster, we intend to explore how, if
 at all, is the LSS modulating galaxy evolution.  Fig.~\ref{type}
 shows the distribution of morphological type of filaments belonging
 to both the cluster samples. The errors were calculated using
 binomial statistics, where, if within a fixed radial range, $n$ out
 of $N$ clusters in a given sample had a particular type of filament
 feeding them, then the likelihood of a cluster to have that class of
 filament is given by $L \propto f^{n}(1-f^{n})^{N-n}$ \citep{d3}. The
 error on the fraction of clusters is given by the standard deviation
\begin{equation}
\Delta f_{x} = \sqrt{\frac{n(N-n)}{N^3}}.
\label{frac-err}
\end{equation}
Each distribution was normalised to unity. The number of filaments
classified for each cluster sample are listed in
Table~\ref{filament-type}.  The choice of the magnitude cut of $M_{r}
\leq -20.5$ excludes all low luminosity galaxies from the analysis,
thus significantly underestimating the density of some of the galaxy
filaments. Hence the morphological class for a large fraction of the
galaxy filaments could not be determined by visual inspection of
density contours on the projected planes. On the other hand this also
implies that the uncertainty in determining the morphological class of
the `clean' filament sample is very small.

Fig.~\ref{type} suggests that the CN clusters are more likely to be
fed by Type II (warped) filaments. The number of filaments associated
with individual clusters do not show any particular trend.
  
\begin{table}
\centering
\caption{Type of filaments} 
 \label{filament-type}
\begin{tabular}{c|c|c|c|c|c|c|c|}
\hline
 Sample   & I   & II  & III  & IV  & V  & undetermined \\ \hline
 SB       & 18  & 10  &  2   &  1  & 7  &  47          \\ \hline
 CN       & 25  & 35  &  6   &  3  & 9  &  68          \\ \hline
\end{tabular}
\end{table} 

 \section{Discussion}
 \label{discussion}

 Superclusters are the most massive and youngest structures in the
 Universe, within which groups and clusters of galaxies evolve through
 interactions and mergers. In this hierarchical model, clusters are
 assembled by accreting surrounding matter in the form of galaxies and
 galaxy groups \citep[\eg][]{b8,struck06,oemler09} through large-scale
 filaments \citep{zes,col2,p2,fadda,p08,edwards10}. On the course of
 their journey from the sparsely populated field to the dense cores of
 clusters, galaxies encounter a range of
 different environments, under the
 influence of which they evolve, not only by exhausting their gas
 content in forming stars, but also by losing the gas to their
 surroundings. Thus, one of the most crucial links required to fully understand
 galaxy evolution is to find how do the galaxies in cluster cores
 become passive while those in the field remain star-forming. 

 \begin{figure}
\centering{
{\epsfig{file=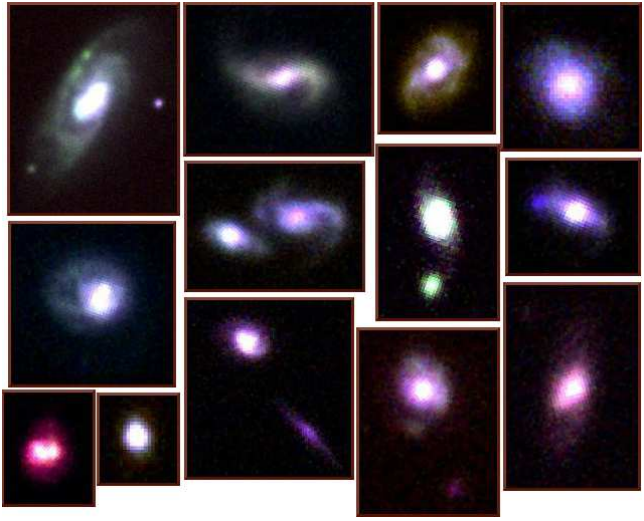,width=8.5cm}}}
\caption{True colour SDSS images of some of the typical \s~galaxies found
 on the outskirts (1-2r$_{200}$) of SB clusters in $g^\prime$, $r^\prime$ and $i^\prime$
 filters (represented by ${\it{red}}$, ${\it{green}}$ and ${\it{blue}}$ colours
 respectively). These galaxies cover a wide range in
 morphologies, and some show structural deformities and presence of close neighbours.
 The star-forming regions are predominantly nuclear, but this
might not be surprising since the star formation rate is based on fibre
spectra. }
\label{images}
\end{figure}

\subsection{Excessive star formation on the outskirts of clusters} 

Studies to understand the evolution of galaxies in various
environments generally involve large-angle sky
surveys \citep[e.g.][]{k3}, or target particular environments
\citep{g1,balogh04b,m4,r2,v1}. Despite the well-known
dominance of passively evolving galaxies in clusters, several
\s~galaxies have been serendipitously discovered near the boundaries of
intermediate redshift clusters
\citep[$z\geq0.2$;][]{keel05,moran05,sato06,marcillac07,fadda,oemler09},
as well as in the local Universe \citep{sun02,g1,cortese07,r1}. An
interesting question then is whether the occurrence of such
\s~galaxies in some clusters is due to chance, or do they constitute
a missing link in understanding galaxy evolution and the assembly of
the cosmic web.

In a study of inter-cluster filaments derived from the 2dFGRS,
\citet{p2} and \citet{p08} found that the mean SFR of galaxies
($z<0.1$) in inter-cluster filaments increases at $\sim 3$\,Mpc from the
 cluster centre. However, due to the unavailability of
requisite data for separating AGN from star-forming galaxies, and of SFRs
derived using sophisticated algorithms\footnote{Porter et
  al. (2007, 2008) used a statistical quantity, the $\eta$-parameter
  from the 2dFGRS database, which is derived from the principal
  component analysis of the spectra as a proxy \citep{eta}.}, their
results had some inherent uncertainties. Their results found support
in observations of extended fields around lensing clusters, where
\citet{moran05}, for example, had found that all of their [OII]
emitting galaxies in CL0024 ($z=0.4$) lie around the virial radius of
the cluster. Considering a simple infall model, \citet{moran05}
deduced that the observed excess in the [OII] emitters at the
cluster's periphery can be explained in a scenario whereby an
early-type galaxy on its infall experiences a burst of star formation
at the virial radius lasting 200~Myr and contributing 1\% of the total
galaxy mass. They attributed the burst of star formation to
interaction of galaxies with shocks in the ICM or the onset of
galaxy-galaxy harassment.
 
 Using the SDSS and 2dFGRS spectroscopic data for galaxies, \citet{balogh04b}
 found a critical local projected density
 of $\Sigma_5 \sim 2$\,Mpc$^{-2}$ in the SFR-density relation characterized by the
 near complete absence of galaxies with large EW(H$\alpha$) at densities greater
 than that (see their figure 3). This critical density is similar to that found near the cluster
 boundary \citep[$\sim\!2-3 h^{2}$\,Mpc$^{-2}$;][]{r2}. 
 
 By analysing a sample of close galaxy pairs in the SDSS, \citet{sa} have suggested that
 galaxy-galaxy interactions are most efficient in inducing a burst of
 star formation when the projected distance between galaxies is $\sim
 100$\,kpc and their relative velocity is $\sim$350\,km s$^{-1}$. Furthermore,
  they found that these conditions are most
 effective in low/intermediate density environments. The conditions
 just outside the cluster boundary are thus well suited for galaxies
 to tidally influence each other.  However, given the complicated
 kinematics involved, galaxies do not have enough time to merge
 together, as they progressively drift towards the cluster core with
 an additional effective infall velocity of $\sim$\,220--900\,km
 s$^{-1}$. For instance, even for a broad range of cluster masses
 $10^{12}\!\leq\!M_{\rm clus}\!\leq\!10^{15}$\,M$_\odot$, \citet{ppl}
 find that the maximum infall velocity occurs in a narrow
 cluster-centric radial range of $2\!<\!r_{max}\!<\!6$\,Mpc. 
 
If the dominant process is galaxy-galaxy harassment in the
infall regions of clusters, successive high-speed encounters between
galaxies will lead to the inflow of gas towards the central regions of
the harassed (mostly low-mass) galaxies \citep[e.g.][]{f98}.
\cite{lake98}, for instance, showed that most of the fuel for star
formation in a harassed galaxy may be driven to the central kpc of a
galaxy within 1--2~Gyr in such interactions, and in many cases in
their simulations, half of the gas mass was transferred to the core of
such a galaxy in $<200$~Myr. This will cause rapid star formation in
the core, and of course, if the galaxy is massive enough to have an
AGN, it might be fuelled as well. In Fig.~8, we present a montage of
some of the ``starburst" galaxies within a radial distance of
$1$-$2\,r_{200}$ of the centres of our SB clusters, where we have used
$g^\prime r^\prime i^\prime$ images from the SDSS archive and produced
``true colour" images. We can see that these galaxies represent a wide
variety of morphologies, though they seem to be predominantly disk
galaxies, and many have close neighbours. What is striking from these
images is that most have concentrated nuclear star formation, which is
indicative of such processes in action. One caveat is that our galaxies
were chosen on the basis of spectra obtained with fibres, so, barring
the farthest galaxies in the sample, many spectra have
preferentially sampled the central regions of these galaxies.
   
 The density of the ICM at the periphery of clusters is small. Hence,
 any cluster related environmental process that could directly
 influence the infalling galaxies can be ruled out as a cause of
 enhancement in the SFR of galaxies \citep[but see][whose simulations
 suggest that under some particular conditions the ICM density on
 cluster outskirts is enough for ram-pressure stripping of infalling
 galaxies]{kapferer09}. The local conditions on the outskirts of
 clusters, i.e. small relative velocities between pairs of galaxies,
 and higher galaxy density with respect to the field, seem to favour
 interactions among galaxies, and may lead to the observed enhancement
 in the SFR of galaxies in these regions. Several other observational
 studies \citep{lewis,balogh04b,moran05,sa,p2,p08} and simulations
 \citep{f98,gb} have also concluded that galaxy-galaxy interactions
 are the most suitable explanation for the observed differences
 between galaxies in and around clusters, and those in the field.
  
 The `blip' near $r_{200}$ in the mean SFR profile of SB clusters
 (Fig.~\ref{sfr}) could be a result of these short-time
 interactions. As a galaxy approaches the core of the cluster into which it is
 falling, it begins to experience the dynamical and tidal influence
 of the residual substructure in the cluster \citep{gb}. We would like
 to stress that the reason such a trend has not been widely noticed
 is that it is very difficult to detect and quantify the enhancement in the
 star formation activity, occurring on such small timescales and confined to
 such narrow regions on sky
 \citep{ppl,p08}. Thus, it is not surprising that despite stacking a fairly
 significant sample of clusters, only a subdued signal was
 detected here.
 
 \begin{figure}
\centering{
{\rotatebox{270}{\epsfig{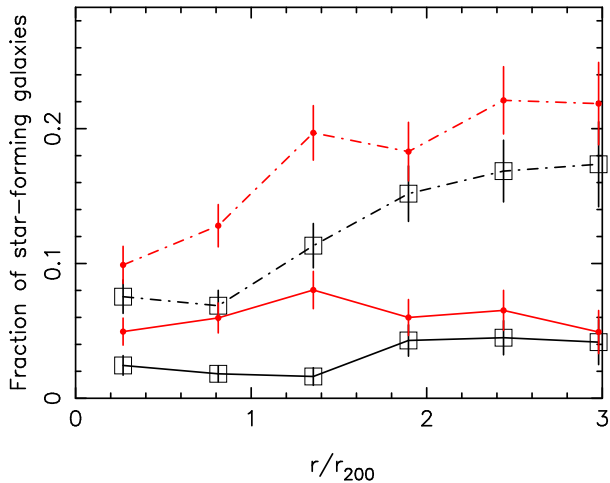}}}}
\caption{Fraction of star-forming galaxies in the SB ({\it{red circles}}) and CN
 ({\it{black squares}}) cluster samples respectively, as a function of scaled cluster-centric
  distance. The distributions shown by {\it{solid}} lines correspond to the selection of
  star-forming galaxies as log SFR/$M^*\geq-10.5 $ yr$^{-1}$ \& SFR $\geq10$ M$_{\odot}$yr$^{-1}$,
  while for the {\it{dot-dashed}} lines the criteria are log SFR/$M^*\geq-10.$ yr$^{-1}$ \& SFR $\geq1$
  M$_{\odot}$yr$^{-1}$, respectively. The fraction of star-forming galaxies in the CN clusters is almost
  always lower than the SB clusters, irrespective of how star-forming galaxies are selected. }
\label{frac-sb}
\end{figure} 
  
 Figure~\ref{frac-sb} provides some more insight into the difference in the SFR profile
 of the SB and CN clusters. A comparison of the fraction of star-forming galaxies
 as a function of the cluster-centric radius for the two samples shows that the fraction of
 starburst galaxies is higher in the SB clusters at almost all radius, irrespective of how the
 star-forming galaxies are selected.
% the fraction in the CN sample, within the cluster radius is almost null.
 This evidently shows
 that the CN clusters are likely to be dynamically relaxed, virialized systems, where the total
 rate of star formation is lower than the SB clusters.
  
 The other peak in the mean SFR profile (Figure~\ref{sfr}), observed nearer the cluster
 centre, could be a consequence of the onset of the cluster-related
 processes such as ram-pressure stripping, or it could
 be the result of an alignment of several
 cluster-feeding filaments along the line-of-sight. While the
 confirmation of latter requires data at other wavelengths and
 sophisticated modelling, the former can only be tested in
 simulations. We leave this issue for later consideration.
 
\citet[][their Fig.~12]{k3} showed that the D$_n$4000 feature in
star-forming galaxies is well correlated with the internal extinction
(measured in SDSS $z$-band).  For D$_n$4000 $\sim$ 1 (most of our
galaxies have D$_n$4000$\gtrsim 1$; see Fig.~\ref{d4000-h}), the
internal extinction can be as high as $1.5$ magnitudes (in the
$z$-band).  Hence it is very likely that the number of \s~galaxies
observed at visible wavelengths may be severely underestimated, making
observations of this phenomenon even more difficult with data in one
wavelength regime alone \citep[also
see][]{g1,mahajan09,wolf09,mahajan10}.
 
A cluster with a deep potential well attracts relatively more
galaxies compared to one with a shallower potential, which implies
 that the increased galaxy density on the outskirts of a cluster implies
 higher probability for galaxy-galaxy interactions in more massive clusters. 
 In the literature, however, there seems to be an
ambiguity in the observational results.  In a study of 25 clusters ($0
< z \leq 0.8$), \citet{pog} found that although the fraction of
emission-line galaxies is anti-correlated with the cluster's
$\sigma_v$ at high redshift, the intermediate mass ($\sigma_{v} \sim$
500-600 km s$^{-1}$) $z \sim 0$ clusters do not show any systematic
trend. \citet{b97} showed that at $z \leq 0.1$ the fraction of
emission-line galaxies is higher by $\sim$ 33\% in clusters with
$\sigma_{v} > 900$\,km s$^{-1}$ as compared to those with $\sigma_{v}
< 600$\,km s$^{-1}$. On the other hand, in a study of 17 clusters from
the 2dFGRS, \citet{lewis} found no difference between the mean
luminosity normalised SFR (proxy for specific star formation rate)
profile of their clusters separated around $\sigma_{v} = 800$\,km
s$^{-1}$. In this paper, using \lx~as a proxy for the mass for almost 70\% of
 our clusters, we do not find any difference in the mass distribution of the
 two samples, but we find the distributions of the $\sigma_v$ to be significantly different.
 We believe this is because $\sigma_v$ represents the dynamical state
 of a cluster, i.e. the degree of relaxation, rather than the mass of the cluster, as is
 often used in the literature. 
   
\begin{figure}
\centering{
{\rotatebox{270}{\epsfig{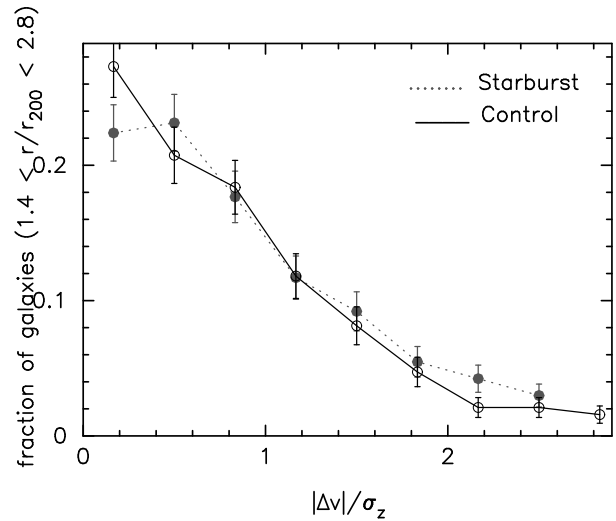}}}}
\caption{Fraction of galaxies ($1.4 \leq$ r/r$_{200} \leq 2.8$)
 in the SB ({\it{doted}}) and CN ({\it{solid}}) cluster samples
 as a function of scaled relative velocity. No significant peak
 at 0 or above implies that these galaxies are a mix of infalling
 and backsplash galaxies \citep{gkg}.}
\label{backsplash}
\end{figure}
 
It might be argued that the observed enhancement in the SFR of
galaxies, on the periphery of clusters, is due to the mixing of
`backsplash' galaxies \citep[\eg][and references
therein]{gkg,mmr} with those falling into the cluster for the
first time. This `backsplash' population comprises galaxies which have
crossed the cluster core at least once and are observed on the other
side of its periphery \citep{b6,mamon04,r2,gkg}. In their simulation,
\citet{gkg} showed that backsplash galaxies can be detected by their
distinct centrally peaked velocity dispersion distribution in the
radial interval 1.4-2.8r$_{200}$, while a non-zero velocity
distribution peak shows that the galaxies are falling into the cluster
for the first time.
 
This hypothesis was tested by plotting the distribution of the
relative line-of-sight (los) velocities of galaxies in the interval
1.4-2.8r$_{200}$. The distribution for SB (and CN) cluster samples, as
shown in Fig.~\ref{backsplash}, showed no prominent peak at $|\Delta v|/\sigma_z = 0$.
 This implies that the galaxies in the infall regions of the cluster samples
studied here are a mix of infall and backsplash populations. This is
in agreement with earlier studies \citep{r2,pk06} confirming that some
emission line galaxies in and around clusters are a result of
backsplash, but a pure backsplash model does not explain all their
properties. Using simple models on similar data, \citet{mmr} have
shown that even a single passage through the cluster core is enough to
completely quench the star formation in a galaxy.
  
Earlier studies showed that SB galaxies in distant ($z \sim 0.5$)
clusters are likely to be low luminosity dwarfs \citep{pog99}.  As
mentioned before, in the observation of enhanced
[OII] emission at 1.8~Mpc from the centre of CL0024 \citep[$z = 0.4$;][]{moran05},
 all of the emission-line galaxies are dwarfs (M$\rm_{B}>-20$). 
 Elsewhere, \citet{hb} found an excess of blue dwarf galaxies
($M_{r}>-18$) around 1.5\,Mpc from the centre of the Shapley
supercluster, while \citet{mahajan10} discovered a similar incidence
of excessive blue dwarfs on the outskirts of the Coma cluster ($z =
0.03$). In the Coma supercluster ($z = 0.03$), dwarf galaxies exhibit
a stronger star formation-density relation than their massive
counterparts \citep{mahajan10}.  In an accompanying study of dwarf
galaxies in the Coma supercluster, \citet{mahajan11} showed that all
the post-starburst k+A galaxies are dwarfs, and are found in or around
the clusters and groups in the supercluster. 

These results are in
agreement with those presented by \citet{smith}, who showed that the
red-sequence dwarf ($M_{r} \gtrsim -17.5$) galaxies on the outskirts
of the Coma cluster have been truncated of star formation.  Due to the
flux limitations of the SDSS data used in this work, the low
luminosity ($M_{r} \geq -20.5$) galaxies could not be included in this
analysis, but the picture portrayed above will hold for them too. The
only exception in case of dwarf galaxies might be the absence of the
second burst seen near ($\sim 0.5 r_{200}$) the cluster centre
(Fig.~\ref{sfr}). This is because amongst others, evaporation of cold
gas becomes very important for dwarf galaxies
\citep{f1,boselli06}. Hence even a relatively softer encounter with a
more massive galaxy may lead to loss of all gas via tidal stripping
and/or evaporation.
 
 \subsection{Star formation and the large-scale structure} 
 \label{lss}
 
 In order to evaluate their role in the evolution of galaxies, it is
 crucial to understand the properties of galaxies on supercluster
 filaments.
High resolution simulations suggest that low over-density regions
 like galaxy filaments can host a significant fraction of gas with temperatures
 between 10$^{5}$-10$^7$ K \citep{co,croft,poc,kd,edwards10b}. This gas may be detected
 as radio emission \citep[\eg][]{kim89}, or through bremsstrahlung emission
 in soft X-rays \citep[\eg][]{kd}. But not only is the signal too weak for the present day detectors,
 the former can be employed only for very nearby cluster pairs.
 Hydrodynamical dark matter (+ gas) simulations \citep[e.g.][]{kd}
 find that due to very low gas densities and low temperatures ($<
 10^7\,$K), it is very difficult to observe these filaments directly.
 Recent Suzaku observations \citep{mit12} seem to have confirmed \citep[also][from ROSAT]{kb99} 
 an excess of hot X-ray gas in the filament
 joining the two rich Abell clusters in the core of the Shapley
 Supercluster.  These observations are currently very difficult, and
 with no advanced X-ray observatory in the horizon, indirect ways of
 understanding the role of filaments in the evolution of cosmos, by
 studying the effect of the filamentary environment on the galaxies
 are necessary \citep[\eg][]{p2,boue08,fadda,p08}.
 
  \begin{figure}
 \centering{
 {\rotatebox{270}{\epsfig{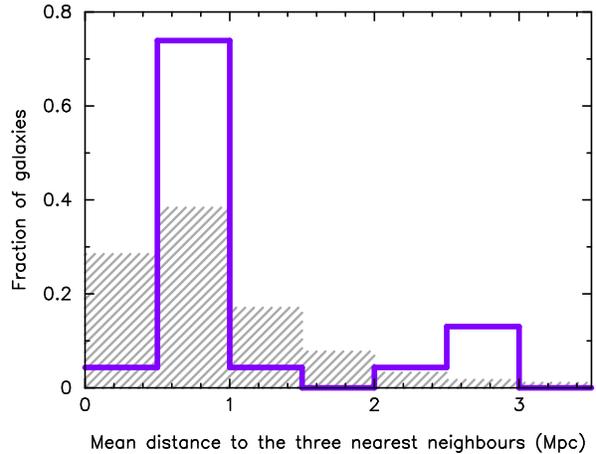}}}}
\caption{The mean distance to the three nearest neighbours, within a
  redshift slice of $\pm 2,000$\,km s$^{-1}$, for galaxies
  1--2\,r$_{200}$ from the cluster centre. The {\it{purple}}
  distribution shows the starburst galaxies, while the
  {\it{grey}} hatched histogram represents other cluster galaxies
  at the same radial distance from the cluster centre. We exclude
  galaxies which do not have at least three neighbours within
  this range.}
 \label{d3}
 \end{figure} 
 
 By employing semi-analytic modelling, \citet{gonzalez08} showed that
 galaxies with similar local density but different `global'
 environment (\ie different positions on the cosmic web), have subtle
 but non-negligible differences in their SFR and colours. They
 attribute this result to the coherent motion of galaxies approaching
 the clusters through large-scale filaments
 \citep{p2,fadda,p08,edwards10}. \citet{struck06} discussed an
 interesting scenario, whereby galaxies in a small galaxy group
 crossing a cluster may be pulled towards each other, thus increasing
 the local galaxy density by an order of magnitude, and the
 probability of galaxy-galaxy interactions by about a factor of 100
 (density squared). Fig.~\ref{d3} provides some evidence in support of
 this scenario \citep[also see][]{g1,moss06,oemler09}.  For galaxies
 at $1\!\leq\!r_{200}\!\leq\!2$ from the cluster centre, the
 mean projected distance to the three nearest neighbours within $\pm
 2,000$\,km s$^{-1}$ of the starburst galaxies is $<1$~Mpc in 78\% of
 the cases. However, for all other cluster galaxies in the same radial
 range, this is true only about 66\% of the time, suggesting that at a
 given cluster-centric distance, starburst galaxies are more likely to
 be in a relatively denser environment. This is worth verifying in
 larger samples as they become available.
 
 Type~I (straight) filaments exist between close ($\lesssim$ 5-10 Mpc)
 cluster pairs \citep{col05,kd,pdh}. However,
the majority ($\geq$ 50\%) of
 the filaments in the cosmic web are of type~II (warped). Some attempts
 have also been made to understand the density profile of type~I
 filaments \citep[also see Edwards et al. 2010b]{col05}.  Their relatively
 shorter length \citep{col05,kd,pdh} and higher matter density may
 imply that the probability of galaxy-galaxy interactions occurring
 amongst galaxies on these filaments is higher than on their
 counterparts. Fig.~\ref{type} favours this scenario. It shows that
 $\sim 32\%$ of the (morphologically classifiable) filaments feeding
 the comparison (CN) clusters are of type~I (straight), as compared
 to $\sim$47\% in the SB clusters. This is also in sync with the
 results of \citet{col05}, who showed that in the cold dark matter
 Universe, more massive structures are more strongly clustered than
 their less massive counterparts. Hence, the unrelaxed SB clusters
 ($\sigma_{v}\gtrsim 500$ km s$^{-1}$) are more likely to
 have a straight filament connecting them to a nearby companion
 cluster. We note that the detection of a significant fraction of our
 filaments remains uncertain because of low galaxy ($M_{r}\leq-20.5$)
 densities, so the quantities mentioned in this section may only be
 considered representative and not precise.
 
 On a final note, we suggest that SB clusters are most likely dynamically
 unrelaxed clusters, presently being assembled via galaxies falling in through
 straight filaments. The occurrence of starburst galaxies on
 the periphery of these clusters is a direct consequence of the way
 local conditions (\ie galaxy density and relative velocity) are affected by the
 global layout of the cosmic web.
  
 \subsection{Future implications}   
     
 Compared to galaxies at $z \sim 0$, distant galaxies are more
 gas-rich. Hence it would not be an exaggeration to say that such
 events of spectacular starbursts would be more commonly seen at
 higher redshifts. Confirmation of this, however, can only result from
 the study of an unbiased list of high redshift clusters, sampled upto
 their infall regions \citep[\eg][also see Coppin et al.  2011 for
 recent observations of star formation related FIR emission on the
 outskirts of clusters at $z \sim 0.25$]{ked,v1}.
 
 Once such data are available, one can ask whether the
 excess of star-forming galaxies observed in clusters at higher
 redshifts, the well known `Butcher-Oemler effect' \citep{bo} is
 linked to the occurrence of starbursts in infalling galaxies, which
 appear to be nearer the cluster centre in projection
 \citep[\eg][]{oemler09}. For a sample of X-ray-luminous clusters at
 $0.18\! <\! z \!<\! 0.55$, 
\citet{e2} showed that the galaxy population in
 the cluster cores are the same at all epochs, but higher redshift
 clusters have a steeper gradient in the fraction of blue galaxies
 when the regions outside the core ($>0.5 r_{500}$) are considered. Such
 colour-based studies, however, may have inherent problems due to the
 non-negligible fraction of galaxies that deviate from the normal
 correspondence between photometric colours and spectroscopically
 derived SFR \citep[see][for instance]{mahajan09}. Another interesting
 idea to explore, though relatively more plausible in simulations, is
 finding the likelihood of detecting  a \s~galaxy on a filament
 feeding a cluster along the line of sight. Such a galaxy will appear
 near the centre of cluster, yet showing physical and intrinsic
 properties as that of its counterparts far away from the cluster core
 \citep[for instance see][]{oemler09}.
 
 It would be very interesting to study the intermediate mass/dwarf
 ($M_r>-20.5$) galaxies, in the infall regions of the SB clusters,
 which may be harassed even more easily by their relatively massive
 counterparts. One needs to perform a detailed morphological analysis of the
 starburst galaxies, such as those studied in this work. Considering
 galaxy-galaxy harassment to be the dominant environmental mechanism
 affecting the star formation properties of galaxies on cluster
 outskirts, one expects to see signatures of tidal interactions with
 neighbouring galaxies in visible or other short wavelength data
 \citep[\eg][]{smith10}. Unfortunately, the spatial resolution of the
 SDSS photometry, used in this work, is not suitable for this, which we
 leave for future follow-up.

\section{Summary and Conclusions}
\label{conclusions}

Using the integrated SFR (and SFR/$M^*$) derived by B04 for galaxies
in the SDSS DR4, the star formation properties of galaxies
($M_r\leq-20.5$) in the nearby ($0.02 \leq z \leq 0.15$) Abell
clusters were analysed. The main results are summarised below:
 
\begin{itemize}
 \item Star formation activity in galaxies on the outskirts of some clusters
 is enhanced near their periphery (r$_{200}$) and further into the clusters at $\sim 0.5$r$_{200}$. 
 In agreement with several other observational and simulations' studies, we
 find that this enhancement in the SFR of galaxies on the clusters' periphery is
 most likely due to enhanced galaxy-galaxy interactions.
 \item Starburst galaxies on cluster outskirts inhabit relatively denser
 environments, compared to other cluster galaxies at similar distances from the
 cluster centre.
 \item Clusters with evidence of enhanced star formation activity on
  their periphery are likely to be dynamically unrelaxed, and fed by
  straight and well-defined supercluster filaments.
 \item The infalling galaxies ($1.4\leq\rm r/r_{200}\leq2.8$)
 in our clusters are a mix of galaxies that have passed through the cluster
 core at least once (the `backsplash' population), and those which are falling
 into the cluster for the first time.
 
\end{itemize}
  
\section{Acknowledgements}

Funding for the Sloan Digital Sky Survey (SDSS) has been provided by
the Alfred P. Sloan Foundation, the Participating Institutions, the
National Aeronautics and Space Administration, the National Science
Foundation, the U.S. Department of Energy, the Japanese
Monbukagakusho, and the Max Planck Society. The SDSS Web site is
http://www.sdss.org/. 

This research has made use of the X-Rays Clusters Database (BAX)
which is operated by the Laboratoire d'Astrophysique de Tarbes-Toulouse (LATT),
under contract with the Centre National d'Etudes Spatiales (CNES).
 Mahajan and Raychaudhury thank Prof. Trevor J. Ponman
 and Prof. Bianca Poggianti for their critique, that helped in improving the
 contents of this paper. We would like to thank the anonymous reviewer whose 
 suggestions helped us in presenting this paper in a broader context. Mahajan was
 supported by grants from the ORSAS, UK, and the University of Birmingham. 

\label{lastpage}

%%%%%%%%%%%%%% Bibiliography %%%%%%%%%%%%%%%%%%%%%%%%%%%%%%%

\end{document}